\documentclass[sigconf,natbib=false]{acmart}
\usepackage{makecell}
\usepackage{enumitem}
\usepackage{subcaption}
\usepackage{tcolorbox}
\tcbset{boxsep=3pt,left=0pt,right=0pt,top=0pt,bottom=0pt}
\RequirePackage[datamodel=acmdatamodel,style=acmnumeric]{biblatex}
\setcopyright{none}

\renewcommand\footnotetextcopyrightpermission[1]{}

\newcommand{\todo}[1]{\textcolor{red}{[TODO] #1}}

\begin{document}
%\title{Navigating Infinity Fabric Interconnect: Towards Efficient Inter-APU Communication in AMD HPC Systems}
%\title{A Deep Dive into Inter-APU Communication Efficiency on AMD MI300A Multi-APU Systems with Infinity Fabric Interconnect}
%\renewcommand{\shorttitle}{A Deep Dive into Inter-APU Communication Efficiency on AMD MI300A Multi-APU Systems}
\title{Inter-APU Communication on AMD MI300A Systems \\via Infinity Fabric: a Deep Dive}
\renewcommand{\shorttitle}{Inter-APU Communication on AMD MI300A Systems}

\author{Gabin Schieffer, Jacob Wahlgren, Ruimin Shi}
\email{{gabins,jacobwah,ruimins}@kth.se}
\affiliation{KTH Royal Institute of Technology \country{Sweden}}
\author{Edgar A. León, Roger Pearce, Maya Gokhale}
\email{{leon,pearce7,gokhale2}@llnl.gov}
\affiliation{Lawrence Livermore National Laboratory \country{USA}}
\author{Ivy Peng}
\email{ivybopeng@kth.se}
\affiliation{KTH Royal Institute of Technology \country{Sweden}}

\acmConference[MEMSYS 2025]{The International Symposium on Memory Systems}{October 7-8, 2025}{Washington DC, NY, USA}
\acmYear{2025}
\copyrightyear{2025}

\renewcommand\footnotetextcopyrightpermission[1]{} % removes footnote with conference information in first column
\pagestyle{plain} % removes running headers

\begin{abstract}
%Large-scale scientific applications, spanning multiple CPU cores and GPUs, require efficient communication between processors. The AMD MI300A Accelerated Processing Unit (APU) integrates a GPU, CPU cores, and high bandwidth memory on a single package. Four APUs connected via Infinity Fabric form a tightly-coupled compute node. To fully exploit this architecture, understanding data movement strategies and programming interfaces is critical. This work classifies data movement approaches and associated programming interfaces. From this classification, we evaluate the achievable performance with a set of benchmarks. Our study covers direct memory access from GPU kernel, explicit data movements, and collective multi-GPU communication. We assess the use of HIP APIs and MPI routines, along with the GPU-specialized RCCL library. Additionally, we adapt two scientific applications to utilize available communication interfaces and allocators and compare measured performance with benchmarking results. Our results highlight key design choices for maximizing performance on multi-APU AMD MI300A systems.

The ever-increasing compute performance of GPU accelerators drives up the need for efficient data movements within HPC applications to sustain performance. Proposed as a solution to alleviate CPU-GPU data movement, AMD MI300A Accelerated Processing Unit (APU) combines CPU, GPU, and high-bandwidth memory (HBM) within a single physical package. Leadership supercomputers, such as El Capitan, group four APUs within a single compute node, using Infinity Fabric Interconnect. In this work, we design specific benchmarks to evaluate direct memory access from the GPU, explicit inter-APU data movement, and collective multi-APU communication. We also compare the efficiency of HIP APIs, MPI routines, and the GPU-specialized RCCL library. Our results highlight key design choices for optimizing inter-APU communication on multi-APU AMD MI300A systems with Infinity Fabric, including programming interfaces, allocators, and data movement. Finally, we optimize two real HPC applications, Quicksilver and CloverLeaf, and evaluate them on a four MI100A APU system. 
%In this context, understanding available data movement strategies and their respective attainable performance is key to optimize communication in HPC applications. 
%Our results highlight key design choices for maximizing performance on multi-APU AMD MI300A systems.

\end{abstract}

\begin{comment}
\begin{CCSXML}
<ccs2012>
<concept>
<concept_id>10010147.10010169</concept_id>
<concept_desc>Computing methodologies~Parallel computing methodologies</concept_desc>
<concept_significance>500</concept_significance>
</concept>
<concept>
<concept_id>10010520.10010521.10010528.10010530</concept_id>
<concept_desc>Computer systems organization~Interconnection architectures</concept_desc>
<concept_significance>500</concept_significance>
</concept>
</ccs2012>
\end{CCSXML}
\ccsdesc[500]{Computing methodologies~Parallel computing methodologies}
\ccsdesc[500]{Computer systems organization~Interconnection architectures}
\end{comment}

\begin{teaserfigure}
    \centering
    \includegraphics[width=0.95\linewidth]{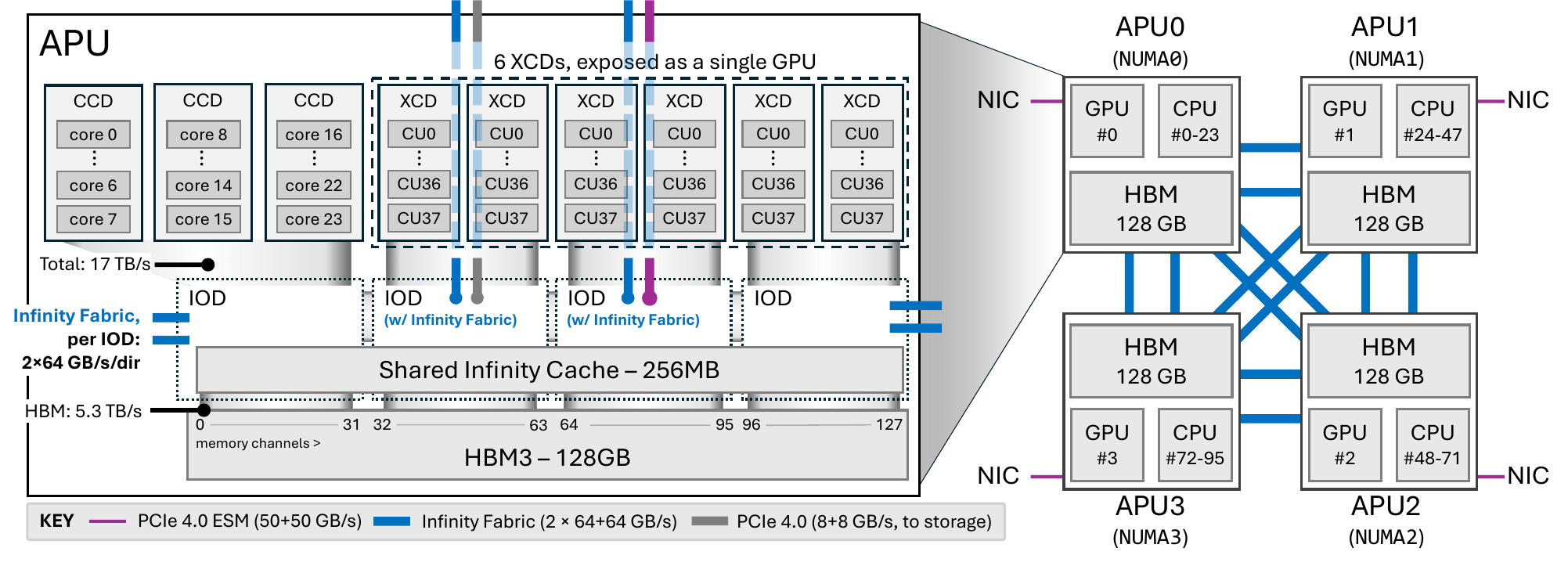}
    \caption{Node Architecture (right) with four MI300A APUs, and detailed APU architecture (left). The Infinity Fabric (in blue) interconnects the four APUs. From the user perspective, each APU is a NUMA node in this cache-coherent NUMA system.}
    \vspace{8pt}
    \label{fig:node-arch}
\end{teaserfigure}

\maketitle

\section{Introduction}
\begin{comment}
\begin{figure*}[bt]
    \centering
    \includegraphics[width=0.95\linewidth]{figures/node-arch-wide.pdf}
    \caption{Node Architecture (right) with four MI300A APUs, and detailed APU architecture (left). The Infinity Fabric (in blue) interconnects the four APUs. From the user perspective, each APU is a NUMA node in this cache-coherent NUMA system.}
    \label{fig:node-arch}
\end{figure*}

\end{comment}

Multi-GPU nodes have become prevalent in leadership supercomputers and High-performance Computing (HPC) systems. Starting from pre-exascale machines, where nodes feature four to six powerful GPUs, a compute node today can be unprecedentedly powerful in providing computing capacity that used to require hundreds and even thousands of CPU-only nodes. Recently, two exascale supercomputers, El Capitan (No.~1) and Frontier (No.~2), have their compute nodes based on multiple AMD GPUs~\cite{loh2023research,atchley2023frontier}. In particular, ORNL's Frontier supercomputer consists of four AMD Instinct MI250X GPUs per node, and two Graphics Compute Dies~(GCD) per GPU, presenting as an eight-GPU node to users. LLNL's El Capitan supercomputer features four AMD MI300A Accelerated Processing Units per node, each APU combines a CPU part and a GPU part on the same package. Although integrated CPU and GPU are not new to the mobile and laptop markets, HPC and data centers mainly use discrete GPUs. In fact, the AMD MI300A APU is the first integrated CPU and GPU that specifically targets HPC. This work provides a timely understanding of data movement mechanisms and strategies for efficient communication on multi-APU systems.  

Even before the emergence of multi-APU nodes, programming and utilizing multi-GPU systems efficiently has been a challenge~\cite{atchley2023frontier,de2024exploring,fusco2024understanding}. In HPC applications, two main programming paradigms are used: the first one uses a single process in the shared-memory model and leverages NUMA-aware thread binding to schedule tasks into multiple GPUs; the second one continues the MPI-based distributed-memory model to have a separate process mapped for each GPU. Efficient data movement on multi-GPU systems has been identified as a critical optimization aspect. For that, recent vendors have introduced high-performance cache-coherent interconnects, such as Nvidia NVLink-C2C and AMD Infinity Fabric, for connecting GPUs~\cite{fusco2024understanding,smith2024realizing}. The complex meshes of GPUs could result in multiple data paths between two communication endpoints, and consequently, different hardware engines along the route may be utilized for acceleration~\cite{cai2021dgcl,choi2022memory}.

This work provides an in-depth understanding of data movement and communication on the emerging multi-APU systems and optimization strategies for preparing HPC applications. Since GPU-accelerated applications are often constrained by how fast the data can be supplied to GPU for computation, efficient APU-APU communication becomes increasingly important for exploiting the full system potential. We base our study on the recently implemented multi-APU node, exemplified by El Capitan supercomputer. %and Tuolumne supercomputers.
The large number of GPU and CPU dies and Infinity Fabric interconnects create a complex node-level mesh~\cite{smith2024realizing}, as illustrated in Figure~\ref{fig:node-arch}. To ensure a representative coverage of the study, we start with a taxonomy of communication mechanisms on multi-APU nodes, classifying them into four categories -- direct GPU kernel access, explicit memory transfer, point-to-point and collective inter-process communication. 

Guided by the taxonomy, we design a set of micro-benchmarks for dissecting and quantifying the performance of data movement using available software interfaces and compare the obtained performance with the peak hardware capacity. In addition to the benchmarking results on MI300A systems, we also provide a comparison with those on MI250X systems. Different from discrete GPUs, on APU, CPU and GPU share a unified memory, and thus the memory management on APU may impact communication. Therefore, we expand the study to evaluate the impact of memory allocation methods and CPU or GPU first-touch strategies on data movement. Moreover, we assess the effectiveness of specialized hardware units, such as SDMA, the XNACK mechanism, and interaction with Linux's Heterogeneous Memory Management (HMM) system. Finally, we demonstrate in two case studies using the Quicksilver and CloverLeaf applications. By optimizing their communication bottlenecks, we achieve up to $2.15\times$ speedup in communication. 

We made the following contributions in this work:
\begin{itemize}[leftmargin=*]
    \item We provide a taxonomy of communication on multi-APU systems, including direct access, explicit transfer, point-to-point, and collective communication;
    \item We propose a methodology for benchmarking communication and the impact of data paths, programming interfaces, and allocators;
    \item We provide the first quantitative characterization of AMD MI300A-based multi-APU systems and identify key optimization insights for efficient APU-APU communication;
    \item We evaluate in two HPC applications Quicksilver and CloverLeaf the effectiveness of inter-APU communication optimization strategies.
\end{itemize}
\section{AMD MI300A based Multi-APU Systems}
Starting with El Capitan % and Tuolumne
supercomputer, multi-APU nodes with integrated CPU and GPU parts become accessible to HPC applications. This section introduces APU, interconnects, and memory management on multi-APU nodes.

\subsection{Accelerated Processing Unit}

{\iffalse
\begin{figure}
    \centering
    %\includegraphics[width=\linewidth]{figures/single-apu.pdf}
    \includegraphics[width=\linewidth]{figures/single-apu-gray.pdf}
    \caption{One MI300A APU is composed of three Core Complex Dies (CCDs), totalling 24 CPU cores; accompanied by eight Accelerator Complex Dies (XCDs), exposed to the user as a single GPU. Each XCD is composed of 38 Compute Units (CUs), totalling 228 CUs per APU.}
    \label{fig:single-apu}
\end{figure}
\fi}

The AMD MI300A is an Accelerated Processing Unit (APU), which combines a CPU and a GPU on the same package, sharing a single physical memory region. More details on its unified physical memory between CPU and GPU are described in~\cite{wahlgren2025iiswc}. This unified physical memory design contrasts with the one taken for the Nividia Grace Hopper Superchip, where two physical memory spaces are still used, but interconnected by cache-coherent NVLink-C2C~\cite{schieffer2024harnessing}. While APUs have been widely used in consumer electronics for a long time, the introduction of such system in HPC is recent. AMD MI300A APU is built on the principle of chiplets. In the manufacturing process, chiplets are design blocks with a well-defined set of functionalities, which can be reused and combined to design more complex hardware. The inset of Figure~\ref{fig:node-arch} highlights the hardware characteristics at a high level. %This section, adapted from~\cite{smith2024realizing}, details the hardware characteristics of an MI300A APU, 

MI300A APUs are composed of a combination of Core Complex Dies (CCDs), which implement CPU cores; Accelerator Complex Dies (XCDs), which form a GPU; memory dies, which use the HBM3 technology; and Input/Output dies, which implement memory-side caching and IO abilities for the attached processors, XCDs and CCDs. On each APU, three CCDs are used; each CCD exposes eight AMD Zen 4 CPU cores, for a total of 24 CPU cores per APU. On the GPU side, the MI300A APU features six XCDs, with 38 compute units (CU) per XCD, totaling 228 compute units over the entire APU. In the simplest configuration, the six XCDs are exposed to the user as a single GPU, with no explicit control of the mapping of GPU kernels to XCDs. The \textit{compute partitioning} feature of the system allows to partition the six XCDs to offer user control on the workload mapping on the XCDs. 

On the GPU side, the compute units implement the CDNA3 microarchitecture~\cite{smith2024realizing}. The L1 data cache is 32~KB per CU, with a cacheline size of 128~bits, L1 instruction cache is 64~KB, shared between pairs of two CUs. The L2 cache is shared between all CUs of a single XCD, with 4MB per XCD. All memory traffic to/from the XCD is coalesced in the L2 cache, where cache-coherence is also enforced with the rest of the APU. On the CPU-side, the CPU cores implement the Zen~4 microarchitecture. The L1 data cache is 32~KB, and the L1 instruction cache is 32~KB. Each CPU core has 1~MB of L2 cache. All cores of a CCD share a 32~MB L3 cache.

A key particularity of this system is the \textit{Infinity Cache}, which is last-level cache (LLC), shared between all XCDs and CCDs, and implemented on the memory side. The entire LLC is 256~MB, distributed into 128 slices of 2~MB each. Each slice is paired with exactly one of the 128 memory channels~\cite{smith2024realizing}. This cache is implemented as part of the Input Output Die (IOD). In total, each APU features 128~GB of HBM3 memory. The physical memory is distributed across eight HBM stacks, with two stacks attached to each IOD. This configuration leads to a total of 512~GB of HBM3 memory over the entire quad-APU compute node.

{\iffalse
----------- DETAILS -------------

Also see https://chipsandcheese.com/p/inside-the-amd-radeon-instinct-mi300as

GPU cache:
	L1-d cache: 32KB (line: 128 bits), per CU
	L1-i cache: 64KB, shared for 2 CUs.
	L2 cache: 4MB, per XCD, coalesce all memory traffic for the XCD, coherence enforced

CPU cache:
	L1-d cache
	L1-i cachie
	L2 cache: 1MB
	L3 cache: 32MB, per CCD

Memory:
	8 stacks of HBM3, two per IOD.
	“4 KB sequential addresses map to the same HBM stack before moving on to another HBM stack based on physical address hashing scheme.”
	128 GB
	128 memory channels
\fi}

\subsection{Infinity Fabric Interconnect}

At a higher level, MI300A HPC systems are built from a four-APU node architecture, where four MI300A are grouped onto a single board composing an HPC node. Notably, this configuration is featured on the El~Capitan supercomputer and is the focus of this work. Figure~\ref{fig:node-arch} presents this architecture.

The key element in this system is the Infinity Fabric (IF) interconnect that connects the four APUs on each node. This interconnect implements the xGMI~3 interface (Inter-chip Global Memory Interconnect~3), also used in other categories of AMD products~\cite{mi300a-sysopt}. A single IF link is 16~bit-wide and operates at a transaction rate of 32~GT/s on the node, as obtained from Linux's sysfs, giving 64~GB/s per direction. Each pair of APUs is connected with two IF links, supporting 128~GB/s bandwidth per direction. This symmetric architecture is depicted in Figure~\ref{fig:node-arch}. This mesh is simpler than those of MI250X-based supercomputers, e.g., Frontier supercomputer, where the Infinity Fabric mesh consists of three tiers of bandwidth between pairs of MI250X GPUs~\cite{de2024exploring}.

For each APU, the IF links are implemented as part of the IODs. Each IOD connects to one IF link, and one configurable link, used for either Infinity Fabric or PCIe~5.0. With four IODs per APU, each APU has a total of six IF links to connect to its peer APUs, with two links dedicated for each peer. In addition, on El~Capitan, on each APU, one x16 PCIe~4.0 ESM (Extended Speed Mode) link connects to the Network Interface Controller~(NIC), with a bandwidth of 50~GB/s per direction. Additionally, on one of the four APUs in the system, a PCIe~4.0 connects the compute node to the near-node storage, with a bandwidth of 8~GB/s per direction.

While the Infinity Fabric interconnect on MI300A system is similar to previous generation on MI250X systems, the mesh drawn by the MI300A Infinity Fabric interconnect is simpler than the previous generation. First, on MI300A system, each APU is directly connected to all peers, whereas on MI250X system, a GPU may need up to two hops to reach other GPUs. In addition, while IF links on MI250X systems have varying bandwidth values for various pairs of GPUs, on MI300A system~\cite{schieffer2024understanding}, all pairs of APUs are connected with the same link bandwidth.

%\subsection{From MI250X to MI300A}
%\todo{to merge with Section 2.2: three tiers } 

% \placeholder{This sections describes the \textbf{hardware} of a multi-APU node}
% \placeholder{Infinity Fabric Interconnect: width, transaction rate, bandwidth, topology, etc.}

{\iffalse
----- DETAILS of the IF Cache -----
Shared Infinity Cache (shared CPU-GPU LLC)
	Memory-side cache, no coherence control.
	Hardware prefetcher
	256MB, per APU
	2MB Inifinity Cache per memory channel (total 256MB)

\fi}

\subsection{Memory Management on Multi-APU Nodes}
From the user perspective, a multi-APU node is a Cache Coherent NUMA (non-Uniform memory Access) system, where each APU is exposed as a NUMA node, grouping the GPU, the 24 CPU cores, and the 128~GB of HBM3 memory. \textit{Node-level memory coherence} is managed transparently for the programmer, so that updates to one APU's memory by any processor (either CPU or GPU) are reflected in all cached copies of the data. Depending on whether the data is accessed by GPU or CPU, memory coherence may be ensured at either hardware or software level. Coherence between each CPU with the rest of the system is achieved through the use of probe filters at hardware level, while coherence between GPU and other GPUs and CPUs in the system is ensured through software support~\cite{smith2024realizing}.

On each APU, CPU and GPU maintain their respective page tables, similar to the previous generation of AMD MI250X GPUs. The GPU page table is distinct from the CPU page table despite sharing the same physical memory space. When the GPU performs a memory operation on a virtual address that is not mapped in the GPU page table, a page fault occurs. In general, a page fault in a GPU kernel terminates the kernel with an error. On AMD MI300A, such a failed memory access will be replayed by leveraging a hardware feature called \textit{XNACK}, which can be enabled by setting the environment variable \texttt{HSA\_XNACK=1}. When XNACK is enabled, together with Linux's Heterogeneous Memory Management (HMM) system, GPU kernels can access system-allocated memory allocated with, e.g., \texttt{malloc}.

{\iffalse
\begin{figure}[ht]
    \centering
    \includegraphics[width=0.9\linewidth]{figures/data_transfer.pdf}
    \caption{Direct in-kernel memory access, performed across Infinity Fabric from a GPU-side compute unit.}
    \label{fig:enter-label}
\end{figure}
\fi}
\section{A Taxonomy of multi-APU Communication}
\label{sec:taxonomy}
\begin{figure}[bt]
    \centering
    \includegraphics[width=\linewidth]{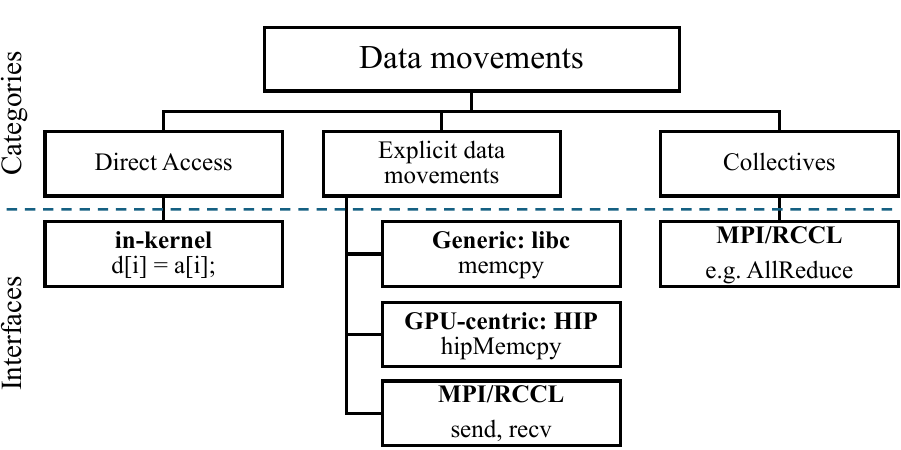}
    \caption{A taxonomy of communication on multi-APU systems, associated data movement categories and programming interfaces and libraries.}
    \label{fig:taxonomy-tree}
\end{figure}

In this section, we present a taxonomy of communication on emerging multi-APU nodes, represented by the El~Capitan supercomputer. Figure~\ref{fig:taxonomy-tree} presents the taxonomy, including their data movement approaches and available programming interfaces.

% \placeholder{Describe the general method: (1) we first establish a classification for transferring data within a multi-GPU application; (2) we extract what APIs/interfaces fall into what category; (3) we evaluate the performance of each API, and identify the pitfalls, and potential issues/limitations. (4) at the light of those issues/limitations, we evaluate real-world applications.}

{\iffalse
\begin{table}
    \centering
    \caption{Programming interfaces grouped by type of access, along with allocation interfaces and first-touch location, for data transfer between APUs on MI300A system.}
    \includegraphics[width=\linewidth]{figures/taxonomy_table.png}
    \label{tab:taxonomy-table}
\end{table}
\fi}

\subsection{Direct Access}
In GPU applications, memory accesses are performed within GPU kernel code, using load and store instructions. Such access offers the lowest latency and highest bandwidth when performed on local GPU memory, physically residing on the GPU where it is accessed from. However, modern GPU hardware and software provide the ability to access data located remotely, either in the host's physical memory or in another GPU's physical memory. Direct access provides the highest level of granularity compared to other data movement approaches, as only the data being accessed is transferred to the accessing processor. This can be beneficial, for example, in applications with complex communication patterns, where the exact extent of the data to be accessed is unknown at runtime, e.g. in graph processing application. However, as the data remain remotely-resident, direct access is not suitable for applications with well-known communication patterns, or performing repeated accesses. 

% Previous generations of GPUs rely on managed memory to achieve such ability, where the GPU driver ensures that the GPU is able to access data when requested. However, modern systems rely on Linux's HMM for this purpose, where the mechanisms are integrated within the Linux kernel.

\subsection{Explicit Data Movement}
Explicit data movement refers to an approach where data are explicitly copied or moved to the memory attached to the processor where it is used. GPU applications are heavily reliant on this principle for several reasons. First, as previous generations of GPUs did not support direct access to remote-located memory, data were necessarily resident in local GPU memory before a GPU kernel could be executed. Second, explicit data movement is advised for performance considerations, as direct data access to remote memory is limited by the available bandwidth of the interconnect, with a theoretical limit 128~GB/s for MI300A, which is dramatically lower than the bandwidth of the local high bandwidth GPU memory, with a theoretical value of 5.6~TB/s on MI300A. In addition, GPUs feature hardware units that are specialized in data copy and do not use the compute capabilities of the Compute Units. These hardware units, referred to as SDMA engines (System Direct Memory Access engines) in AMD's terminology, can perform copy operations in parallel with kernel execution. Therefore, explicit data movement offers the opportunity to overlap communication and computation at a high level and for large memory regions instead of relying solely on the instruction pipelining abilities of GPU compute units. % latency-hiding abilities of GPU compute units. %load-store instruction when used in direct in-kernel data accesses.

\subsubsection{C standard library's memcpy}
The C standard library provides a memory copy function, memcpy, which performs a copy between two buffers. The actual implementation of memcpy is compiler-dependent. While the GNU C Compiler implements memcpy as a while loop performing a series of C load-and-assign operations, Clang's implementation of memcpy is platform-dependent.

\subsubsection{GPU-centric APIs}
HIP~\cite{hip} runtime exposes APIs to perform explicit data movement, namely hipMemcpy, originally designed for host-to-device and device-to-host data movement, but now supports any type of data movement, and hipMemcpyPeer, which is dedicated to inter-GPU peer-to-peer data movement. On AMD hardware, hipMemcpyPeer is a thin wrapper around hipMemcpy~\cite{hipMemcpy_impl}. We note here that HIP APIs rely on lower-level APIs,  HSA~\cite{hsa} memory management APIs, to execute actual copy operations. Directly using such lower-level APIs offers the opportunity for hardware-tailored performance optimizations. However, as their portability is limited and their use in real-world applications remains marginal, they fall out of the scope of this work.

\subsubsection{Multi-process Point-to-Point Communication}
The use of Message Passing Interface (MPI) is ubiquitous in HPC applications, to distribute computations across several processes. The MPI standard defines routines to send and receive data across processes, \texttt{MPI\_Send} and \texttt{MPI\_Recv}, respectively. From the application perspective, those routines are semantically equivalent to an explicit data copy operated with, e.g., memcpy. A notable difference is that instead of copying data between buffers allocated by the same process, data are transferred between buffers belonging to different processes, located on distinct processes on potentially distinct compute nodes. Naturally, such operation is more complex than for intra-process explicit data movement.

While MPI is the \textit{de facto} standard in HPC, AMD's ROCm Collective Communication Library (RCCL), which is a collective communication library specialized for GPU communication, also provides point-to-point communication routines, similar to the ones defined in the MPI standard. For single-process applications, RCCL can be utilized by itself without any other dependency, while for multi-process applications, RCCL is used in conjunction with MPI.

\subsection{Collective Communication}
Collective operations are a category of communication that involves all communicating endpoints. Collective operations often consist of several communication processes. Such a communication pattern is heavily relied upon in HPC and distributed machine learning workloads. While collective communication routines are underlyingly implemented with a series of point-to-point communications, they also include computations on the collaborating processors. In HPC, the use of MPI is ubiquitous for collective communication. However, the RCCL library also appears as a strong alternative, which offers comparable capabilities, and is specialized for GPU-GPU communications.
\section{Methodology}
In this section, we describe the benchmarking design for characterizing each communication category in the taxonomy in Section~\ref{sec:taxonomy}. We also introduce two real-world applications used for case study and the testbed environments. 

\subsection{Benchmark Design}
\subsubsection{Direct Access}
To evaluate the performance of direct data access on quad-MI300A system, we use a GPU variant of the STREAM benchmark, where two buffers are allocated and initialized in one APU's memory, using the hipMalloc allocator. A GPU kernel is then executed on another APU. This kernel reads data from and stores data to the buffers allocated on the peer APU. By measuring the bandwidth of the copy operation, we can evaluate the achievable copy bandwidth over an APU-APU Infinity Fabric link.

To evaluate the unidirectional bandwidth of the Infinity Fabric link between two APUs, we rely on the hipMemcpy API. In the default configuration, this API uses dedicated GPU hardware units to perform a copy operation and does not rely on GPU kernels. By setting the environment variable \verb|HSA_ENABLE_SDMA=0|, we override this behaviour and force the hipMemcpy API to use highly optimized copy kernels, referred to as ``\textit{blit}'' kernels, which directly implement a copy operation using load-store instructions executed on the GPU's compute units~\cite{blit_kernel}, instead of relying on dedicated hardware units. We use this approach to measure the peer-to-peer copy bandwidth, achievable with direct data access. In addition, with a pointer-chasing approach we measure the latency of local memory accesses and remote memory accesses over Infinity Fabric.

\subsubsection{Explicit Data Movements}
We develop a bandwidth measurement benchmark to evaluate the performance of explicit data movement APIs. We construct our benchmark with Google's benchmark framework library~\cite{googlebenchmark}. We define three phases: allocation, first-touch, and data movement. The benchmark measures the bandwidth of the copy operation in the data movement phase; a warm-up phase is included, and measurements are repeated 10 times. The benchmark allows changing the underlying interface used in each of the three phases. For the allocation phase, four allocators can be used: the system allocator \texttt{malloc}, the HIP GPU-centric allocator \texttt{hipMalloc}, the HIP managed memory allocator \texttt{hipMallocManaged}, and the HIP host-memory allocator. The second phase, first-touch, refers to the initialization, which can be done either by a CPU thread or the GPU. For CPU first-touch, we use libc's \texttt{memset} function. The data movement phase uses either the \texttt{memcpy} function or HIP \texttt{hipMemcpy} API call. We observed that both \texttt{hipMemset} and \texttt{hipMemcpy} fail with "invalid argument" when called on a non-HIP allocated buffer. For \texttt{hipMemset}, we work around the issue by implementing a simple GPU kernel to initialize the memory. For \texttt{hipMemcpy}, registering the malloc-allocated memory with \texttt{hipHostRegister} allows calling \texttt{hipMemcpy} on the allocation. To control the location of the source and destination buffers, we set the locality of CPU threads with \texttt{numa\_run\_on\_node}, which constrains a CPU thread to execute on a specific NUMA node, that is, APU. For HIP-related APIs, we ensure execution on the desired GPU by using \texttt{hipSetDevice}.

\subsubsection{MPI/RCCL Point-to-Point Communication}
We use the OSU micro-benchmark suite (OMB)~\cite{omb} to evaluate the bandwidth and latency of point-to-point send and receive operations. We compare MPI routines, widely used in HPC applications, with routines provided as part of ROCm Communication Collectives Library (RCCL), which are specialized for GPU-GPU communication. Underlyingly, these routines use system-specific APIs, similar to hipMemcpy or memcpy, to perform the actual data movement. We use the benchmarks \texttt{osu\_bw} and \texttt{osu\_lat} for MPI, \texttt{osu\_xccl\_bw} and \texttt{osu\_xccl\_lat} for RCCL. The latency benchmarks execute a ping-pong latency measurement. The bandwidth benchmark initiates a series of fixed-size back-to-back MPI messages with \verb|MPI_ISend| from a sender process and receives those messages on another process using matching \verb|MPI_Recv| operations. The wall-clock time of 10,000 \verb|MPI_ISend| and the corresponding \verb|MPI_Recv| operations are measured.

The MPI implementation in use, Cray MPICH, dynamically changes its underlying communication paths depending on message sizes, e.g., it uses shared memory CPU buffers for intra-node communication of messages no larger than 1024 bytes and uses SDMA-accelerated direct peer-to-peer GPU communication for larger messages. Therefore, we use two configurations for MPI. First, we enforce direct peer-to-peer GPU-GPU inter-process communication by setting \verb|MPICH_GPU_IPC_THRESHOLD| to \verb|0|, denoted as \textit{GPU direct}. Second, we enable CPU staging by setting \verb|MPICH_GPU_IPC_ENABLED| to \verb|0|, denoted as \textit{CPU staging}. In addition, we ensure that GPU-aware capabilities are enabled in the MPI implementation by setting \verb|MPICH_GPU_SUPPORT_ENABLED| to~\verb|1|. We further evaluate several combinations of allocators for the source and destination buffers to evaluate the ability of the MPI implementation to map copy operations to actual hardware capabilities, in various circumstances.

% We further adapt the allocator for the source and destination buffers in the bandwidth benchmark to study the impact of memory allocators on MPI and RCCL communication. While the MPI specification abstractly defines point-to-point routines, the MPI implementation must be able to efficiently map these operations to the actual hardware capabilities. In particular, in the context of regular GPU-accelerated systems, the MPI implementation must be able to distinguish between CPU-located and GPU-located memory and adjust the copy mechanism to use accordingly. On MI300A, this aspect is slightly shifted, as on one APU, a single physical memory space is shared between CPU cores and GPU. Our system's MPI implementation, Cray MPICH supports GPU-aware communication. We ensure that this feature is enabled by setting the environment variable \verb|MPICH_GPU_SUPPORT_ENABLED=1|.

% Furthermore, for small message sizes, the MPI implementation relies on CPU-staging of buffers to perform inter-process data movements, e.g., in \texttt{MPI\_Send} and \texttt{MPI\_Recv} routines. The threshold is set by the system administrator and is 1024 by default on our system. This is controlled by the environment variable \texttt{MPICH\_GPU\_IPC\_THRESHOLD}.

\subsubsection{MPI/RCCL Collective Communications}
We measure the latency of common collective operations using both the RCCL and MPI benchmarks provided as part of the OSU micro-benchmark suite~\cite{omb}, for various message sizes and numbers of GPUs. %In parallel, we use a simple model of collective communication latency, derived from measured latency in point-to-point routines, which we compare with our measurements.

\subsection{HPC Applications}
%Our benchmarking method aims at providing practitioners with a picture of available data movement interfaces and associated achievable bandwidth for inter-APU communication. However, those findings might not necessarily generalize to end applications, which exhibit complex and highly-optimized data movement patterns combined with computations, which are absent from our benchmark approach. To fill this gap, 
We select two real-world GPU-accelerated HPC applications -- \textit{Quicksilver}~\cite{quicksilver} and \textit{Cloverleaf}~\cite{mcintosh2014performance}, for the use case study. Quicksilver is a dynamic Monte Carlo particle transport code that represents the Mercury workload, this workload exhibits unbalanced communication and irregular access pattern. Cloverleaf is a Lagrangian-Eulerian hydrodynamics application, which exhibits a balanced communication pattern with regular access pattern. Both applications have GPU kernels implemented in HIP and rely on MPI for inter-process communication. To demonstrate the effectiveness of communication optimization on multi-APU systems, we adapt their allocation sites and communication interfaces only, and compare them with the original version. In our evaluation, Quicksilver used the CORAL2 Problems 1 and 2 with 2M to 42M particles while CloverLeaf used the bm2028\_short problem with 61440$\times$30720 cells. 
\begin{comment}
Table~\ref{tab:benchmarks} presents the applications and the test cases used for evaluation. 
\begin{table}[ht]
    \centering
    \caption{List of evaluated applications and test cases.}
    \begin{tabular}{c|c|c}
        \hline
        Application & Test case & Problem size \\
        \hline\hline
        Quicksilver & CORAL2-Problem1* & 2M/21M/42M particles \\
        \hline
        Quicksilver & CORAL2-Problem2* & 2M/14M/27M particles \\
        \hline
        CloverLeaf & bm2028\_short & 61440$\times$30720 cells \\
        \hline
        \multicolumn{3}{r}{*shortened to 10 cycles}
    \end{tabular}
    \label{tab:benchmarks}
\end{table}
\end{comment}

\subsection{Testbed}
\begin{table}[bt]
\centering
\caption{Main node characteristics of testbeds.}
\resizebox{\linewidth}{!}{% <------ Don't forget this %    
\begin{tabular}{c|l|l}
                & MI300A Testbed & MI250X Testbed\\
    \hline\hline
    NUMA domains &4 MI300A  &4 MI250X\\\hline
    CPU& 24-core AMD Zen 4& 64-core AMD Trento EPYC\\
    \hline
    GPU& 6 XCDs exposed as 1 GPU & 2 GCDs exposed as 2 GPUs \\
    \hline
    Infinity Fabric& 512 Gb/s links & 288 Gb/s and 400 Gb/s links \\\hline
    Memory & 128~GB HBM3 & 128~GB DDR4, 128~GB HBM2\\
    \hline
\end{tabular}% <------ Don't forget this %
}
\label{tab:system}
\end{table}

%We use the Tuolomne and Tioga supercomputers at Livermore Computing as main testbed. 
We use two testbeds in our study, namely an MI300A system as our main testbed and an MI250X testbed for comparison. Table~\ref{tab:system} summarizes their main node characteristics. On the MI300A testbed, each node is equipped with four AMD MI300A APUs. We use the Cray Programming Environment~24.11 and the \texttt{hipcc} compilation toolchain from ROCm~6.2.1. For point-to-point and collective communication experiments, we use RCCL~2.20.5 and Cray MPICH~8.1.31 (based on ANL MPICH~8.4a2).
%The Flux scheduler is used on these systems and we disable the CPU affinity by setting MPIbind. Applications are executed with one MPI process per APU.

{\iffalse
\subsubsection{Software Stack}
\todo{detail on version}
The following specifications and runtime components are used in MI300A systems. In this list, each component interacts directly with its immediately-following component:
\begin{itemize}
    \item Heterogeneous-computing Interface for Portability (HIP): specification of a kernel language and runtime APIs;
    \item AMD Compute Language Runtimes (CLR): AMD-specific implementation of the HIP specification;
    \item ROCR Runtime and AMD Heterogeneous System Architecture (HSA): runtime library, notably handles kernel launch, memory management, and synchronization;
    \item ROCK Kernel Driver: kernel driver for AMD GPUs.
\end{itemize}
\fi}

\section{Multi-APU Single-Process Communication}

\subsection{Direct Kernel-level Access}
\begin{figure}[bt]
    \centering
    \includegraphics[width=\linewidth]{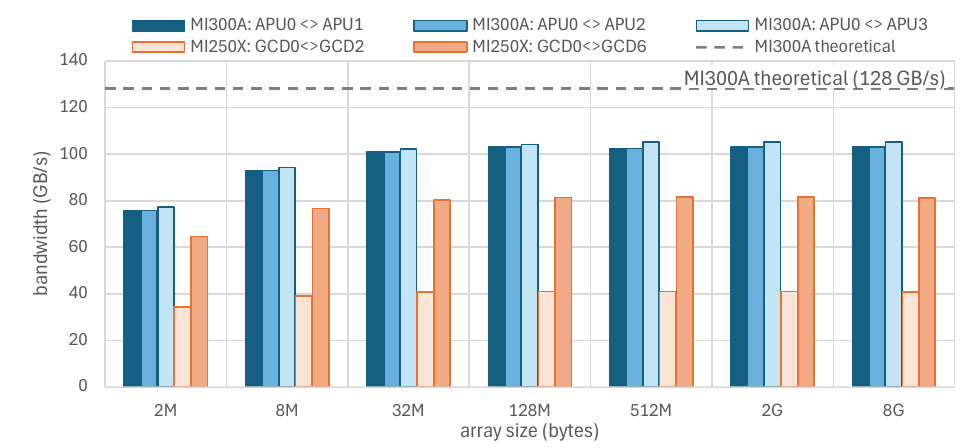}
    \caption{Bidirectional direct access bandwidth obtained with the STREAM Copy kernel, executed on APU0 with data placed on peer APUs. Results on AMD MI250X are obtained by executing on GCD0 with data placed on neighbor GCDs.}
    \label{fig:stream_process_apu0}
\end{figure}

We use a GPU-variant of the STREAM benchmark to evaluate the performance of in-kernel direct remote access. In this benchmark, the arrays are allocated on one APU, using hipMalloc, and a GPU copy kernel is executed on another APU, reading from a peer GPU, and writing back to it. Figure~\ref{fig:stream_process_apu0} presents the results of the copy kernel of the GPU variant of the STREAM benchmark, with a kernel executed on APU0 and data located on APU1, APU2, or APU3, for array sizes from 2~MB to 8~GB. Across all data placements, we observe a bandwidth of 103-104~GB/s, which homogeneous results are consistent with the node topology, where all APUs are directly connected to all other APUs with the same link bandwidth. The measured bandwidth represents 81\% of the theoretical bandwidth of the Infinity Fabric link. We compare the results with the same benchmark executed on MI250X GPUs, with two data placements, based on the non-balanced node topology on MI250X, where Infinity Fabric bandwidth is 50~GB/s for GCD0-GCD2, and 100~GB/s for GCD0-GCD6. In this situation, we reach 82\% and 81\% of the theoretical link bandwidth for GCD0-GCD2 and GCD0-GCD6, respectively. The values of link utilization are similar to values obtained on the newer MI300A.

{\iffalse
\begin{figure}[bt]
    \centering
    \includegraphics[width=\linewidth]{figures/lat_gpu.pdf}
    \caption{GPU memory access latency, measured with a pointer-chasing approach, for data located locally, or on a neighbour APU.}
    \label{fig:lat-gpu}
\end{figure}
\begin{figure}[bt]
    \centering
    \includegraphics[width=\linewidth]{figures/lat_cpu.pdf}
    \caption{CPU memory access latency, measured with a pointer-chasing approach, for data either located locally or on a neighbour APU.}
    \label{fig:lat-cpu}
\end{figure}
\fi}

\begin{figure}
    \centering
    \includegraphics[width=\linewidth]{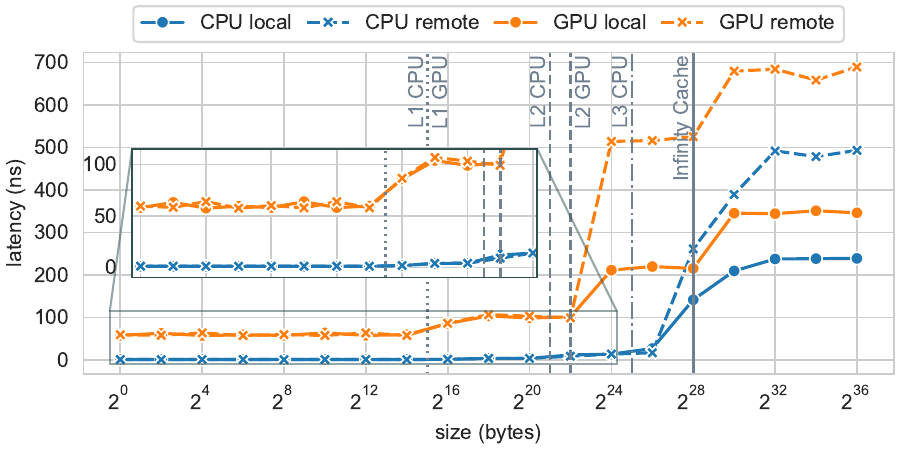}
    \caption{CPU and GPU latency of direct access to local memory, or remote memory (dashed line), located on a neighboring APU. Cache sizes are represented as vertical lines. }
    \label{fig:lat-cpu-gpu}
\vspace{-10pt}
\end{figure}

We measure the latency of direct access to local and remote memory using a pointer-chasing approach, adapted from Google's multi-chase benchmark. For local access, memory is allocated with hipMalloc on the same APU as the pointer-chasing kernel; for remote access, memory is allocated with hipMalloc on a neighbour APU to the one executing the pointer-chasing kernel. Figure~\ref{fig:lat-cpu-gpu} presents the results of this pointer-chasing approach for CPU and GPU, with increasing data size. The latency for local access to HBM memory is 240~ns for CPU, and 346~ns for GPU. For remote data access, the latency increases to 500~ns for CPU access, and 690~ns for GPU access.

\begin{tcolorbox}
\textbf{Observation 1}: Direct GPU kernel access to local and remote APU's HBM has 500~ns and 690~ns latency, higher than CPU's direct access of 240~ns and 346~ns latency. GPU kernel can directly access data on remote APU at 103~GB/s. 
\end{tcolorbox}

\subsection{Explicit Data Movement}
\label{sec:direct-transfer}

In this section, we evaluate the performance of explicit data movement APIs. For this purpose, we compare hipMemcpy and memcpy operations. On the previous generation of AMD GPU, AMD MI250X, the SDMA engines were documented to be unable to fully utilize the Infinity Fabric link bandwidth due to their initial design being optimized for communication on PCIe speeds~\cite{rocm-doc-gpu-mem}.

\subsubsection{Low Transfer Sizes}

\begin{figure}[bt]
    \centering
    \includegraphics[width=\linewidth]{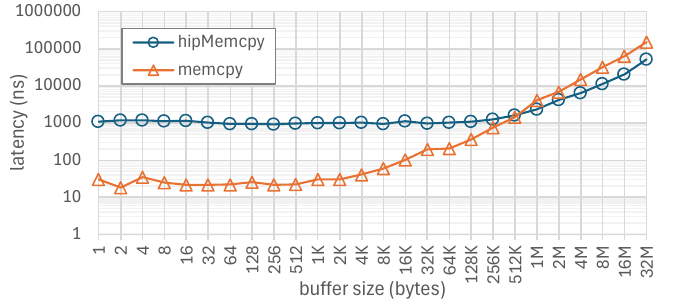}
    \caption{Latency of hipMemcpy and memcpy for an APU-APU transfer on hipMalloc-allocated buffers with CPU first-touch.}
    \label{fig:lat-hipmalloc-memcpy_vs_hipMemcpy}
\end{figure}

For hipMalloc-allocated memory, Figure~\ref{fig:lat-hipmalloc-memcpy_vs_hipMemcpy} presents the latency of memcpy and hipMemcpy operations for low transfer sizes. We observe that memcpy outperforms hipMemcpy for low transfer sizes, up to 512~KB. This is due to the nature of memcpy, which is implemented as a series of load and store instructions, which can operate on cache levels of the system. Therefore, the measured latency is below 100~ns for transfers up to 16~KB. In contrast, hipMemcpy operations are more complex, as they are delegated to the Heterogeneous System Architecture (HSA) runtime, resulting in higher latency. For transfer sizes between 1~byte and 128~KB, a hipMemcpy call represents 1~$\mu$s.

\begin{tcolorbox}
\textbf{Observation 2}: For transfer sizes below 512~KB, memcpy exhibit lower latency compared to hipMemcpy, due to its ability to leverage the various cache levels in the system. 
\end{tcolorbox}

\subsubsection{CPU-side memcpy}
We measure the bandwidth of CPU-side memcpy operation to copy large buffer from APU0 to APU1, we evaluate various allocators, both the system allocator malloc, and the HIP allocators hipMalloc, hipHostMalloc, and hipMallocManaged. We use the compiler-implemented memcpy.

In this experiment, we ensure that physical memory is allocated to both source and destination buffers. For this purpose, we initialize both buffers with an arbitrary value. For the CPU-side first-touch, we use memset, which performs initialization of each of the buffer's elements within a loop. For GPU-side initialization, the hipMemset API cannot be called on a memory region untracked by the GPU driver, e.g., allocated with malloc, resulting in an invalid argument error; instead, we use a GPU kernel for this purpose. % THIS IS REPEATED IN METHOD

\begin{figure}[bt]
    \centering
    \includegraphics[width=\linewidth]{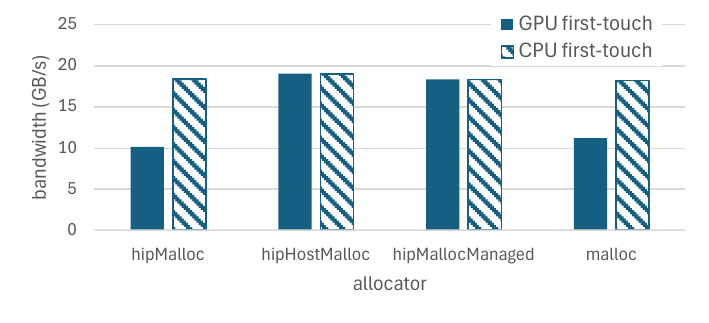}
    \caption{The impact of allocators and first-touch on the maximum bandwidth (GB/s) achieved by memcpy.}
    \label{fig:table_p2p_memcpy}
\end{figure}

Figure~\ref{fig:table_p2p_memcpy} presents the achieved bandwidth for memcpy, when one thread performs a copy operation from APU0 to APU1, with a buffer size of 8~GB. For all allocators and first-touch locations, the copy bandwidth is below 20~GB/s. We suggest that this low bandwidth compared to the theoretical limit of 128~GB/s is due to the nature of the memcpy implementation, which relies on a loop to copy memory from the source buffer to the destination buffer, using load and store instructions. This implementation only leverages one CPU core and, therefore, cannot utilize the full bandwidth offered by the link between APUs. For hipMalloc and malloc allocators, with GPU first-touch, copy bandwidth is significantly lower than for other allocator/first-touch combinations, on the order of 10~GB/s.  % \todo{Hypothesis: GPU first-touch only populates in GPU page table, memcpy then triggers page faults? but then since we have a warmup, we should not see a difference...}

\subsubsection{GPU-centric hipMemcpy}
The hipMemcpy API, provided as part of the ROCm runtime, is designed for data copy in a heterogeneous CPU-GPU system. We measure the copy bandwidth for a data copy from APU0 to APU1, performed with hipMemcpy between two 8~GB buffers. We use the same initialization strategy as for memcpy, where either CPU or GPU performs the first-touch.

For hipMemcpy, data movement is performed using System Direct Memory Access (SDMA) engines by default, which are hardware units for copying data across the system and bypassing compute units, enabling overlap of the copy operation with kernel execution. On MI250X GPUs systems, however, the use of SDMA engines causes under-utilization of the GPU-GPU link.

\begin{figure}[bt]
    \centering

    \includegraphics[width=\linewidth]{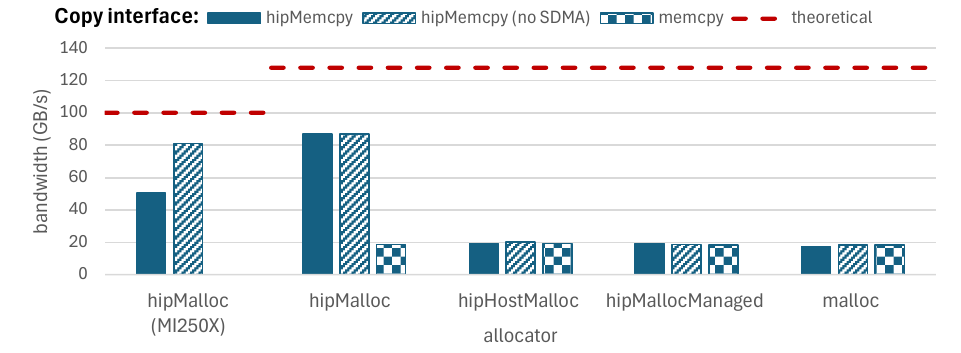}
    \caption{The achieved explicit data copy bandwidth from using hipMemcpy or memcpy for copying data between MI300A APUs.}
    \label{fig:results_p2p_hipmemcpy}
\end{figure}

Figure~\ref{fig:results_p2p_hipmemcpy} presents the achieved bandwidth for explicit data copy between two APUs with hipMemcpy. We observe that hipMemcpy only exhibits the highest copy bandwidth for hipMalloc-allocated buffers, with 90~GB/s. Other allocators are not able to fully leverage the bandwidth of the link and only reach values of bandwidth comparable to those obtained with memcpy. Upon inspection of the hipMemcpy implementation code~\cite{hipMemcpy_impl}, we suggest that these copy operations fall back on standard memcpy calls, executed as single-threaded CPU-side copies. 

In addition, we do not observe any significant difference in bandwidth comparing copy operations using SDMA engines, which is the default behavior, or using direct GPU copy kernels by explicitly disabling SDMA engines. This contrasts with previous generations of AMD GPUs, embodied by the MI250X GPU, where copy operations that rely on SDMA engines are not able to fully leverage the available Infinity Fabric bandwidth. This is due to the SDMA engines on this generation being tuned for PCIe speeds and, therefore, cannot leverage the full link bandwidth offered by Infinity Fabric~\cite{rocm-doc-gpu-mem}. Our results demonstrate that this limitation has been lifted on AMD MI300A, where copy operations using SDMA engines can reach the same level of bandwidth as for direct copy kernels.

{\iffalse 
\begin{tcolorbox}
\textbf{Observation 2}: When accessing data between two APUs, \verb|memcpy| has the lowest latency for small data while \verb|hipMemcpy| can achieve the highest bandwidth of 90~GB/s for data larger than 512~KB, 4$\times$ that of \verb|memcpy|. 
\end{tcolorbox}
\fi}

\begin{tcolorbox}
\textbf{Observation 3}: For inter-APU copy operations above 512~KB, hipMemcpy outperforms memcpy, due to its ability to offload the operation to SDMA engines or GPU copy kernels, thereby enabling the use of the full Infinity Fabric bandwidth.
\end{tcolorbox}

\begin{comment}
\begin{tcolorbox}
\textbf{Observation 3: } Memory allocator and first-touch may impact data transfer bandwidth between APUs. GPU first-touch of \verb|hipMalloc|- or \verb|malloc|-allocated data reduce \verb|memcpy| bandwidth by 50\%.
\end{tcolorbox}
\end{comment}

\section{Multi-APU Multi-Process Communication}
\label{sec:mpi-rccl-p2p}
In this section, we compare point-to-point and collective communication routines between MPI and RCCL. Moreover, we study the impact of different memory allocators on leveraging the Infinity Fabric link bandwidth.

\subsection{Point-to-Point Communication}
\label{sec:p2p-subsec}

\subsubsection{Latency}

\begin{figure}[bt]
    \centering
    \includegraphics[width=\linewidth]{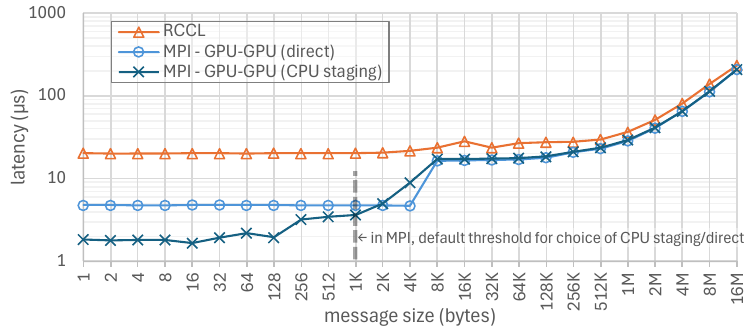}
    \caption{The latency of MPI and RCCL for point-to-point GPU-GPU communication at increased message sizes.}
    \label{fig:lat_mpi_rccl_d2d}
\end{figure}

Our latency measurement results indicate that MPI point-to-point routines with CPU staging achieve the lowest latency. Figure~\ref{fig:lat_mpi_rccl_d2d} presents the inter-APU ping-pong latency measured on hipMalloc-allocated communication buffers at various message sizes. For small message size below 128 bytes, MPI routines with CPU staging have a latency as low as 1.9$~\mu$s, while direct peer-to-peer MPI communication exhibits a 4.8$~\mu$s latency. In contrast, the latency of RCCL is significantly higher than that of MPI for small messages, with a lowest latency of 20~$\mu$s, that is, $10\times$ higher than MPI routines. For direct MPI GPU-GPU communication, we observe a jump in the measured latency when increasing message size from 4~KB to 8~KB. We suggest that this jump indicates a change of behavior in the MPI implementation for messages above 4~KB. However, due to the proprietary nature of the implementation, we were not able to pinpoint the exact cause. Overall, compared to the latency of direct GPU kernel access and \verb|memcpy| in Section~\ref{sec:direct-transfer}, point-to-point communication has a significantly higher latency for small messages. This increased latency is induced by the complex nature of a point-to-point operation, where not only data must be copied, but also expensive inter-process communication is performed.

\subsubsection{Bandwidth}

\begin{figure}[bt]
    \centering
    \includegraphics[width=\linewidth]{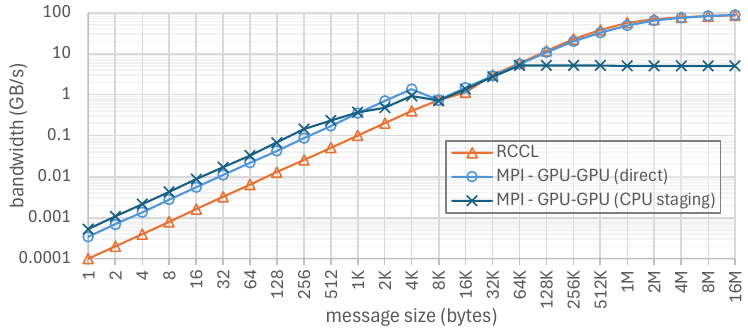}
    \caption{The measured bandwidth of MPI direct GPU-GPU communication, MPI GPU-GPU communication with CPU staging, and RCCL. Destination and source buffers are allocated with hipMalloc.}
    \label{fig:bw_mpi_rccl_d2d}
\end{figure}

% ---  bandwidth ---
Figure~\ref{fig:bw_mpi_rccl_d2d} presents the bandwidth of point-to-point routines in MPI and RCCL with hipMalloc-allocated buffers between APUs. We observe that for message sizes above 8~KB, the bandwidth of RCCL matches the one obtained with direct GPU-GPU MPI communication. We observe that the CPU staging option in MPI outperforms the peer-to-peer GPU-GPU communication for message sizes of 1024~KB or smaller. This is due to the overhead of requesting transfer with SDMA engines in the case of direct peer-to-peer GPU-GPU communication, compared to the low overhead of performing a CPU-side copy between two APUs. As shown in Figure~\ref{fig:bw_mpi_rccl_d2d}, RCCL point-to-point communication routines achieved a maximum bandwidth of 88~GB/s, which is comparable to the bandwidth measured with hipMemcpy APIs.

{\iffalse
\begin{table}[ht]
    \centering
    \caption{Maximum achieved bandwidth (GB/s) of a series of back-to-back MPI send operations, measured with OSU micro-benchmarks. SDMA enabled.}
    \begin{tabular}{c||c|c}
         & \multicolumn{2}{c}{destination buffer} \\
         source buffer & malloc & hipMalloc \\
         \hline\hline
         malloc & 12.5 & 12.3 \\
         \hline
         hipMalloc & 58.2 & 87.9 
    \end{tabular}
    \label{tab:mpi_bw_sdma}
\end{table}
\fi}

\subsubsection{Impact of Memory Allocators}

\begin{figure}[bt]
\centering
\includegraphics[width=\linewidth]{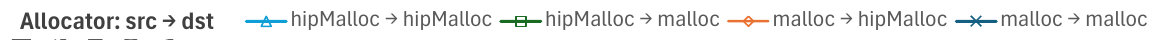}
\begin{subfigure}[t]{0.5\linewidth}
\centering
    \includegraphics[width=\linewidth]{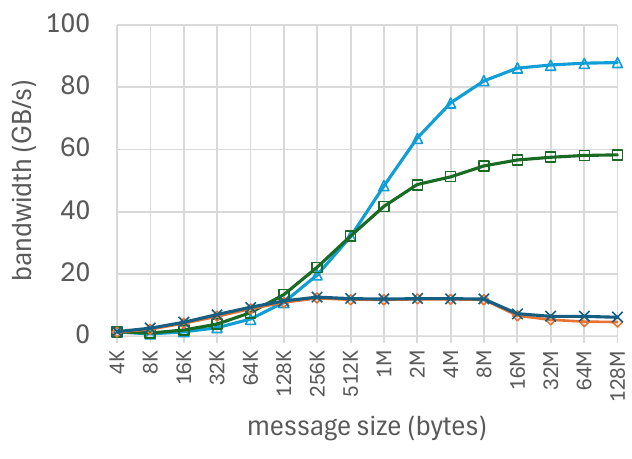}
    \caption{MPI}
    \label{fig:osubw_sdma}
\end{subfigure}
~ 
\begin{subfigure}[t]{0.5\linewidth}
\centering
    \includegraphics[width=\linewidth]{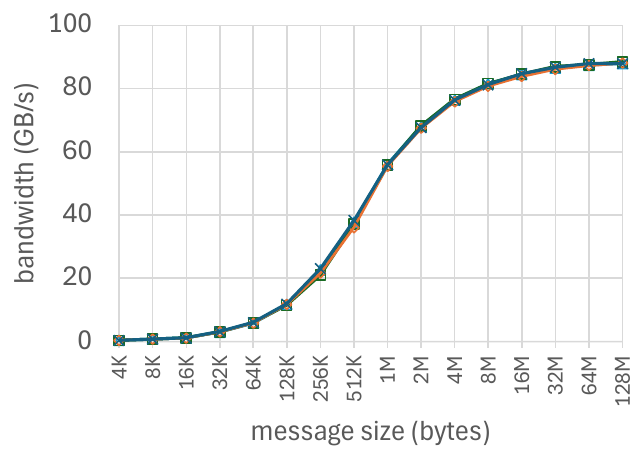}
    \caption{RCCL}
    \label{fig:osubw_xccl}
\end{subfigure}
\caption{The impact of different allocators for the source and destination buffers on point-to-point bandwidth with SDMA enabled.}
\end{figure}

We evaluate the impact of the allocator on point-to-point bandwidth in MPI and RCCL. For MPI, Figure~\ref{fig:osubw_sdma} shows that when the source buffer is allocated with malloc, the maximum bandwidth measured for MPI send/receive operation is 11.7~GB/s. This is comparable to the values obtained with a single-threaded memcpy operation, presented in Section~\ref{sec:direct-transfer}. When both buffers are allocated with hipMalloc, the measured bandwidth is 82~GB/s, which matches the values reported in Section~\ref{sec:direct-transfer}, obtained with hipMemcpy. This indicates that the GPU-aware MPI implementation can efficiently leverage the available inter-APU bandwidth, when using hipMalloc for both source and destination buffers. In contrast, when the source buffer is allocated with hipMalloc and the destination buffer is allocated with malloc, the bandwidth drops to 54~GB/s. We hypothesize that in such scenario, both GPU-only and system page tables are involved in the copy operation. This causes significant overhead. For RCCL, the bandwidth measured under point-to-point routines, presented in Figure~\ref{fig:osubw_xccl}, appears to be insensitive to the choice of allocator. This highlights the ability of RCCL to map the execution of point-to-point routines to the most efficient hardware interface.

\begin{tcolorbox}
\textbf{Observation 4}: RCCL point-to-point routines can efficiently leverage the full Infinity Fabric bandwidth, independent of the choice of allocator. MPI point-to-point routines only achieve the full Infinity Fabric bandwidth when both source and destination buffers are allocated with hipMalloc.
\end{tcolorbox}

% \placeholder{MPI Send/Recv on malloc-allocated buffers doesnt use hipMemcpy.}
% \placeholder{MAIN PROBLEM OVERALL: malloc-allocated buffers are not detected by MPI as able to leverage SDMA engines ---> bad performance}

\subsubsection{The impact of SDMA engines}

\begin{figure}[bt]
    \centering
    \includegraphics[height=3cm]{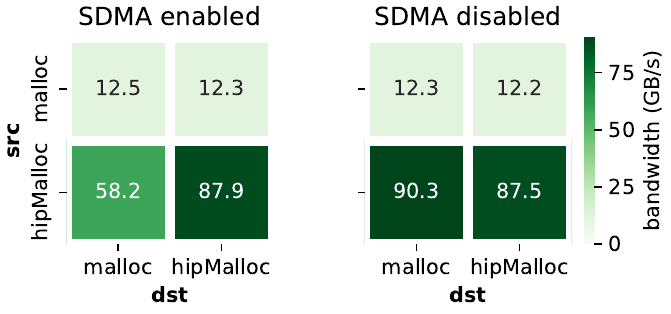}
    \caption{The peak MPI p2p bandwidth using different allocators for source and destination buffers and SDMA settings.}
    \label{fig:mpi_bw}
\end{figure}

We further evaluate the impact of SDMA engines on the bandwidth of MPI send/receive operations between two APUs. Our results in Section~\ref{sec:direct-transfer} demonstrate that SDMA engines on MI300A APUs can fully utilize the Infinity Fabric link. Therefore, for MPI and RCCL, we expect similar bandwidth when disabling SDMA engines. We set the environment variable \verb|HSA_ENABLE_SDMA=0| to measure the bandwidth with disabled SDMA engines. When SDMA is disabled, data movement relying on the HSA runtime will use direct GPU-executed copy kernels to perform data copy. We present the measured bandwidth in Figure~\ref{fig:mpi_bw}, when using either hipMalloc or malloc to allocate source and destination buffers.

In MPI, when the source buffer is allocated with malloc, the state of SDMA engines does not impact the bandwidth, measured at 12~GB/s. This is expected, as the copy mechanism in this situation appears to not rely on GPU's HSA runtime but instead on CPU-side mempcy, which never relies on SDMA engines. When the source buffer is allocated with hipMalloc, and the destination buffer is allocated with malloc, the bandwidth measured with SDMA engines disabled is 90.3~GB/s. This is significantly higher than the 58.2~GB/s bandwidth measured with SDMA engines enabled. For RCCL, we conduct the identical measurements. Our results indicate that SDMA engine state has little impact on RCCL point-to-point bandwidth.

\begin{figure}[bt]
    \centering
    \includegraphics[width=\linewidth]{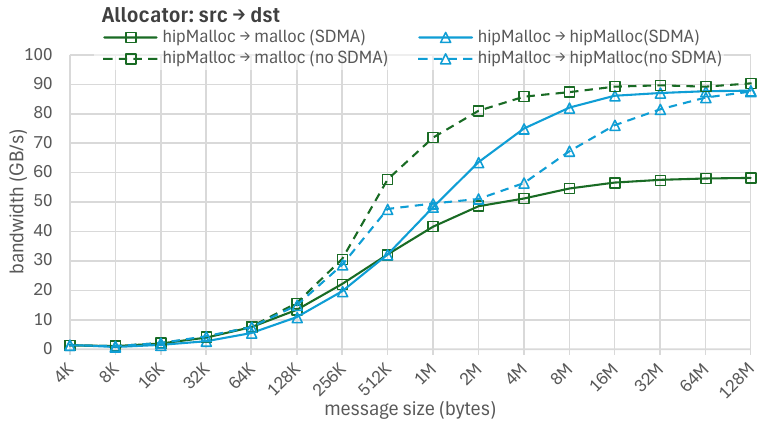}
    \caption{Bandwidth of MPI back-to-back send operations, measured with OSU micro-benchmark, for various allocators for source and destination buffers. Dashed lines indicate that SDMA engines are explicitly disabled.}
    \label{fig:compare_sdma}
\end{figure}

{\iffalse
\begin{table}[]
    \centering
    \caption{With SDMA engines disabled, maximum achieved bandwidth (GB/s) of a series of back-to-back MPI send operations, measured with OSU micro-benchmarks.\todo{replaced by Figure~\ref{fig:mpi_bw}}}
    \begin{tabular}{c||c|c}
         & \multicolumn{2}{c}{destination buffer} \\
         source buffer & malloc & hipMalloc \\
         \hline\hline
         malloc & 12.3 & 12.2 \\
         \hline
         hipMalloc & 90.3 & 87.5 
    \end{tabular}
    \label{tab:mpi_bw_nosdma}
\end{table}
\fi}

Figure~\ref{fig:compare_sdma} presents the point-to-point bandwidth measured with different SDMA settings. The results for a source buffer allocated with malloc are omitted, as they were not influenced by the state of SDMA engines in our experiments. When copying from a hipMalloc-allocated buffer to a malloc-allocated buffer, disabling SDMA engines brings a significant bandwidth improvement for all message sizes. For the largest message size, the bandwidth increases from 58.2~GB/s to 90.3~GB/s. When both source and destination buffers are allocated with hipMalloc, the bandwidth evolution exhibits a different pattern. For message sizes below 1~MB, disabling SDMA engines achieves a higher bandwidth than with SDMA enabled. In contrast, above 1~MB, disabling SDMA has a detrimental effect on bandwidth. However, at the largest message size, the same bandwidth is observed for either SDMA state.
%In detail, with SDMA disabled and hipMalloc for both buffers, the bandwidth increases linearly with the message size, up to 512~KB. Above 512~KB, the increase is linear up to the maximum bandwidth, with a lower slope. 

\begin{tcolorbox}
\textbf{Observation 5}: Bandwidth of MPI point-to-point routines depends on SDMA engine status and allocator choice, with hipMalloc yielding highest bandwidth; RCCL fully utilizes Infinity Fabric link in all evaluated circumstances.
\end{tcolorbox}

\begin{figure*}[bt]
    \centering
    \begin{subfigure}[t]{0.5\linewidth}
    \centering
        \includegraphics[width=\linewidth]{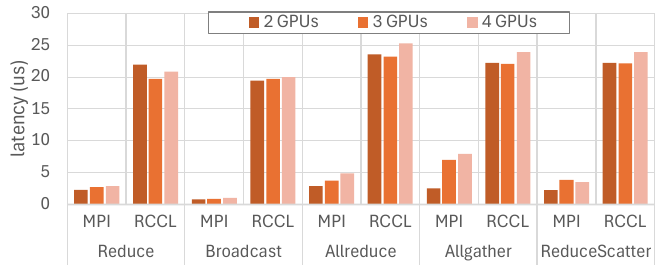}
        \caption{4 bytes}
        \label{fig:collective_lat_osu_xccl_4b}
    \end{subfigure}
    ~ 
    \begin{subfigure}[t]{0.5\linewidth}
    \centering
        \includegraphics[width=\linewidth]{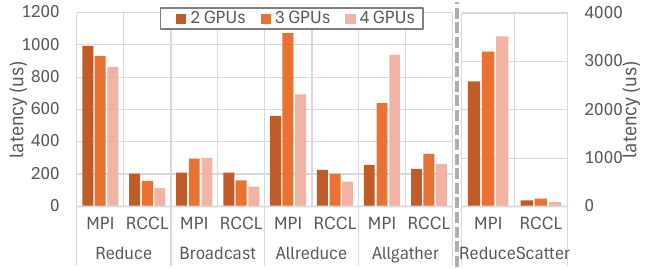}
        \caption{16~MB}
        \label{fig:collective_lat_osu_xccl_16MB}
    \end{subfigure}
    \caption{The measured latency of MPI and RCCL collective operations for 4~bytes and 16~MB messages, with 2 to 4 participating MI300 APUs. ReduceScatter for 16~MB messages uses a separate y-axis. }
    \label{fig:collective_lat_osu_xccl}
\end{figure*}

\subsection{Collective Communication}

\begin{figure}[bt]
    \centering
    \includegraphics[width=\linewidth]{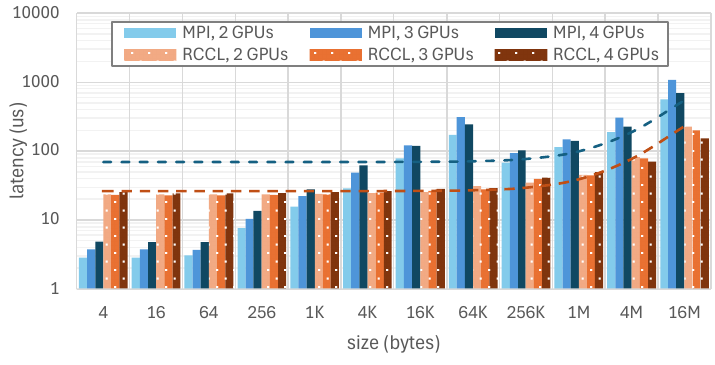}
    \caption{Latency of AllReduce collective operation, with 2 to 4 APUs in the same node participating in the collective. Linear trends are plotted as dashed lines.}
    \label{fig:lat_allreduce}
\end{figure}

We first investigate the scalability of one widely-used collective operation, AllReduce at increased message sizes, in Figure~\ref{fig:lat_allreduce}. %Results are presented in Figure~\ref{fig:lat_allreduce} for 2 to 4 APUs on the same compute node, both for MPI and RCCL collectives. 
For all message sizes up to 4~KB, MPI outperforms RCCL. However, beyond 4~KB, RCCL routines start to exhibit lower latency than MPI. Moreover, RCCL latency scales linearly with the message size, as plotted in the dashed line in Figure~\ref{fig:lat_allreduce}. In MPI, such a linear trend is only observed above 256~KB. This indicates that the runtime of MPI is less predictable than RCCL, likely due a change in the underlying communication interface used by the MPI implementation, depending on buffer size. For instance, by default MPI uses CPU-staging for messages smaller than 1024 bytes, as discussed in Section~\ref{sec:p2p-subsec}.

%In particular, we suggest that low latency values observed for lower buffer sizes are due to the use of CPU-staging in MPI for buffer sizes below 1024 bytes, which, as demonstrated in Section~\ref{sec:p2p-subsec}, achieve significantly lower latency than RCCL.

{\iffalse
\begin{figure*}
    \centering
    \includegraphics[height=4.6cm]{figures/collective_lat_osu_xccl.pdf}
    \includegraphics[height=4.6cm]{figures/collective_lat_osu_xccl_reducescatter.pdf}
    \caption{Latency of collective operations, with various number of partners (APUs) contributing to the collective, from 2 to 4 in the same node. We use a buffer size of 16~MB. \todo{add other message sizes from 1byte - 16 MB}}
    \label{fig:collective_lat_osu_xccl}
\end{figure*}
\fi}

{\iffalse
\subsubsection{Modelling Collective Communication}
We propose a model to derive the latency of collective operations in MPI or RCCL from the size of the collective buffers, the number of participating processors, and the type of collectives. Any collective operation falls within one of three types: one-to-all, where one processor sends data to all other processors (broadcast); all-to-one, where one processor receives data from all other processors (gather); and all-to-all, where all processors send data to and receive data from all other processors (AllGather, AllReduce, and Reduce-Scatter).
\fi}

\begin{comment}
\begin{table}[]
    \centering
    \begin{tabular}{c|c||c|c}
        \hline
        Category & Operations & MPI & RCCL \\
        \hline\hline
        One-to-all & Broadcast & \\
        \hline
        All-to-one & Reduce &  \\
        \hline
                   & AllGather & \\
        All-to-all & AllReduce & \\
                   & Reduce-Scatter & \\
        \hline
    \end{tabular}
    \caption{Category of collective operations, and associated lower latency bound for a 16~MB collective operation, calculated for MPI and RCCL, based on point-to-point MPI communication results.\todo{Shall we remove this table?}}
    \label{tab:model_collectives}
\end{table}
\end{comment}

%We measure the latency of collective operations using OSU micro-benchmark for both MPI and RCCL collectives. 
Figure~\ref{fig:collective_lat_osu_xccl} presents the latency of RCCL and MPI collectives with two to four APUs participating in the collective. %Figure~\ref{fig:collective_lat_osu_xccl_4b} presents the minimal size of 4~bytes, and Figure~\ref{fig:collective_lat_osu_xccl_16MB} presents the largest evaluated size, 16~MB. 
We make the same observation as for AllReduce, where for small messages (Figure~\ref{fig:collective_lat_osu_xccl_4b}), MPI outperforms RCCL. This is expected, as we observed that RCCL point-to-point communication routines exhibited a baseline \textasciitilde$20~\mu$s ping-pong latency. %, which is also visible in those collective operations. 
For larger messages (e.g., 16~MB in Figure~\ref{fig:collective_lat_osu_xccl_16MB}), RCCL collectives outperform MPI routines. As demonstrated earlier, RCCL implementations can leverage the bandwidth of Infinity Fabric links more efficiently than MPI, resulting in higher performance for bandwidth-bound communication, e.g., large messages. %in the default configuration.
ReduceScatter is commonly used in distributed machine learning workloads. For large message sizes, RCCL exhibits a significant advantage of 20-38$\times$ speedup over MPI ReduceScatter.

\begin{tcolorbox}
\textbf{Observation 6}: For messages larger than 4~KB, RCCL collectives lead to $5-38\times$ lower latency than MPI. For messages smaller than 1024 bytes, MPI collectives have the lowest latency.
\end{tcolorbox}
\section{HPC Applications}
\label{sec:app}
In this section, we present two case studies in real-world HPC applications to demonstrate the strategy of optimizing multi-APU communication as identified according to our characterization study. 

\subsection{QuickSilver}

QuickSilver is a multi-process MPI application for dynamic Monte Carlo particle transport problems. As a Monte Carlo code, it uses a large number of particles in simulations, and these particles are spread across the whole domain, which is divided across MPI processes. Thus, communicating particles across processes becomes one time-consuming task of the application. For this purpose, a class \verb|MC_Particle_Buffer| is used, to hold information on particle buffers, and to expose methods to control the exchange of particle data across processors. In the original version, these particle communication buffers are allocated using the system allocator \verb|malloc| and are exchanged using MPI point-to-point routine \verb|MPI_Isend|.

Our profiling results of QuickSilver communication pattern indicate that many small messages are used for communication. As identified in Section~\ref{sec:mpi-rccl-p2p}, RCCL point-to-point communication routines have higher latency than MPI routines for small messages, thus, they will not be used for point-to-point communication. Also, from the characterization, disabled SDMA has positive impacts on the bandwidth of MPI point-to-point communication on all message sizes. Therefore, we disable the SDMA settings for optimization. For experiments, we compile Quicksilver to produce a binary supporting any XNACK state, and change the environment variable \verb|HSA_XNACK| at runtime. Finally, we adapt the allocator for communication buffers from \verb|malloc| to \verb|hipMalloc| because our characterization study indicated that malloc-allocated buffers do not reach the maximum link bandwidth in MPI point-to-point routines. In Quicksilver, this is achieved by adapting the \verb|Allocate| method of \verb|MC_Particle_Buffer|.

%\subsubsection{Communication Time}
% 2 - When HSA_XNACK=0, the allocator *does* matter in the communication time, with hipMalloc being consistenly faster for communication, but not kernel execution.
\begin{comment}
\begin{figure}[bt]
    \centering
    \includegraphics[width=\linewidth]{figures/qs_periter_comm_vs_particles.pdf}
    \caption{The measured communication time in Quicksilver using two CORAL2 input problems at increased number of particles.}
    \label{fig:qs_comm_perit_alloc}
\end{figure}
\end{comment}

% 1 - With HSA_XNACK=1, the choice of allocator does not change runtime.
\begin{figure}[bt]
    \centering
    \includegraphics[width=\linewidth]{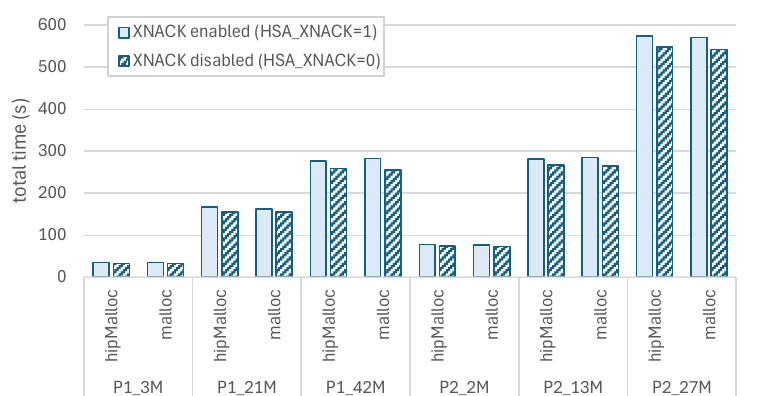}
    \caption{End-to-end runtime measured in Quicksilver for all input problems, comparing the impact of XNACK settings and allocators.}
    \label{fig:e2e_qs_xnack_on_off}
\end{figure}

Figure~\ref{fig:e2e_qs_xnack_on_off} details the runtime of the six Quicksilver test cases, evaluated with XNACK enabled or disabled, using either hipMalloc or malloc for the allocation of buffers used in point-to-point MPI routines. With XNACK enabled, for any fixed problem, the runtime in Quicksilver is insensitive to the selected allocator for allocating communication buffers. However, when XNACK is disabled, we observe the speedup from 5\% to 11\% on the end-to-end execution time. This confirms that the selected optimization is effective for further running Quicksilver simulations on multi-APU systems.

We demonstrate in Section~\ref{sec:mpi-rccl-p2p} that only hipMalloc-allocated buffers can be communicated at full link bandwidth when using MPI. However, when replacing malloc by hipMalloc, the communication time reported by Quicksilver decreases only for the largest test case, Problem~1 with 42M particles, from 6.8~s down to 5.9~s. The reason is that for low transfer sizes, the benefit of using hipMalloc-allocated buffers for communication is outweighed by the overhead of transferring data to those buffers before the actual communication.

%Figure~\ref{fig:qs_comm_perit_alloc} presents the communication time per time step in Quicksilver using our optimized version, as compared to the original version. The $x$-axis represents the number of particle used in the Monte Carlo simulation, which correlates to the communication size. Two input problems P1 and P2 are used, and the communication time is reported as the average of all communication phases in the application. In this test, XNACK is disabled. The results show that communication time scales linearly with the number of particles, which is expected given the problem specifications. When the number of particles used in the Monte Carlo simulation scales from 1.7 million to 42 millions, the gap between the optimized version and the baseline version increases.
{\iffalse
\begin{figure}
    \centering
    \includegraphics[width=\linewidth]{figures/qs_comm_perit_alloc.pdf}
    \caption{Communication time in Quicksilver, reported as an average over all instance of the communication phase in the application. For each of the two problems P1 and P2, three variants are evaluated, with various number of particles. The buffer used in MPI send/receive operations are allocated with either malloc or hipMalloc. XNACK is disabled.}
    \label{fig:qs_comm_perit_alloc}
\end{figure}
\fi}

\subsection{CloverLeaf}

Cloverleaf is a Lagrangian-Eulerian hydrodynamics application. Solvers in such applications heavily rely on send and receive operations. This application exhibits balanced communication, with regular memory access pattern. The baseline HIP implementation supports managed memory, allocated with hipMallocManaged. We preserve MPI calls for smaller message communication operations, such as process synchronization. We add support for system-allocated memory, where the allocation with malloc must be combined with hipHostRegister, to allow hipMemcpy operations. We adapt to use RCCL point-to-point routines for send and receive operations, implemented as part of the \verb|clover_exchange| function. The application is compiled with generic XNACK support. However, executing the RCCL implementation with XNACK disabled caused the program to exit due to errors in the RCCL internal code. Thus, the results are obtained with XNACK enabled.
{\iffalse

\begin{figure}[bt]
    \centering
    \includegraphics[width=\linewidth]{figures/clover_e2e_runtime.pdf}
    \caption{End-to-end runtime (in seconds) of CloverLeaf using the original implementation and our implementation using various memory allocators. All adapted versions are marked with ``*''. Average over five runs.}
    \label{fig:cloverleaf_runtime}
\end{figure}
\begin{figure}[bt]
    \centering
    \includegraphics[width=\linewidth]{figures/cloverleaf_comm_time_with_speedup.pdf}
    \caption{The communication time (in seconds) and speedup ($\times$) in CloverLeaf, for various allocators, comparing the baseline implementation and our implementation.}
    \label{fig:clover-comm-time}
\end{figure}
\fi}

\begin{figure}[bt]
    \centering
    \begin{subfigure}[t]{0.5\linewidth}
    \centering
        \includegraphics[width=\linewidth]{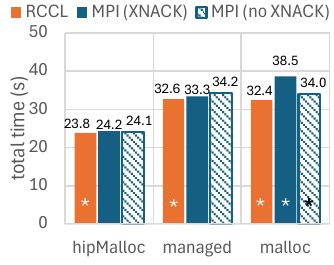}
        \caption{Total time}
        \label{fig:cloverleaf_runtime}
    \end{subfigure}
    ~ 
    \begin{subfigure}[t]{0.5\linewidth}
    \centering
        \includegraphics[width=\linewidth]{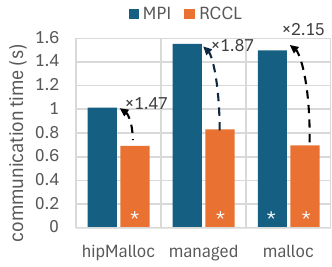}
        \caption{Communication time}
        \label{fig:clover-comm-time}
    \end{subfigure}
    \caption{The total runtime (\subref{fig:cloverleaf_runtime}) and communication time (\subref{fig:clover-comm-time}) of CloverLeaf. ``*'' indicates our adapted versions, ``$\times$'' indicates speedup over MPI.}
    %\label{fig:clover-comm-time}
\end{figure}

Figure~\ref{fig:cloverleaf_runtime} presents the end-to-end runtime of CloverLeaf using MPI or RCCL communication interface and various allocators. For the original implementation, the default hipMalloc-allocated version provides the lowest end-to-end runtime compared to hipMallocManaged and the malloc system allocator. This is consistent with our benchmarking results, presented in Section~\ref{sec:direct-transfer}, which show that for MPI, only hipMalloc-allocated buffers can be exchanged between APUs at high bandwidth. We show that the original MPI version has 15\% higher runtime when using malloc+hipHostRegister, compared to the hipMallocManaged version.

%We demonstrated how the end-to-end runtime is sensitive to the choice of allocators and the communication interface used. However, such observation cannot be solely tied to the performance of communication interfaces in use since only a limited part of the end-to-end runtime is spent in communication. 
To understand how communication optimizations affect the end-to-end runtime, we utilize the internal CloverLeaf timers to quantify improvements in communication time. These timers measure the execution time of the MPI Halo Exchange function (\verb|clover_exchange|), which implements the core of the application's data movement. Figure~\ref{fig:clover-comm-time} presents the communication time in Cloverleaf, for the three evaluated allocators. The reported time is averaged over five trails. We observe that the communication time in the MPI implementation is highly sensitive to the allocator, with values of 1.01~s for hipMalloc, 1.50~s for malloc, and 1.55~s for hipMallocManaged. This is consistent with our characterization results for MPI explicit data transfers. However, the 0.54~s communication time difference observed between the best-performing hipMalloc and hipMallocManaged in the MPI version cannot by itself explain the 9.1~s difference in end-to-end runtime. This observation highlights that while the choice of allocators impacts the communication time between APUs, other factors must be considered, including GPU kernel performance and performance of copy operations within a single APU's physical memory space, with e.g., hipMemcpy or memcpy. For RCCL, the communication time exhibits limited variability across the three allocators, with 0.69~s for hipMalloc and malloc, and 0.83~s for hipMallocManaged. This is also consistent with our benchmarking results and demonstrates how RCCL can achieve efficient communication, with limited impact from the allocator used for the communicated buffers.

In all tested cases in CloverLeaf, the communication-optimized implementation outperforms the original version, with a $1.5\times$ speedup for hipMalloc, $1.9\times$ for hipMallocManaged, and $2.2\times$ for malloc. These results highlight the inefficiency of using non-hipMalloc buffers for MPI point-to-point communication routines. Furthermore, this demonstrates how the use of RCCL enables developers to opt for best-performing allocator for their respective use case or depending on application-specific constraints, while still achieving the highest level of bandwidth for inter-APU communication. %These results confirm, in a real-world application, our findings that RCCL is able to perform efficient data transfer, no matter which allocator is in use. These results suggest that RCCL provides practitioners with the ability to choose an allocator that is best-suited for e.g., in-kernel computations, while achieving high inter-APU communication bandwidth, which neither HIP APIs nor MPI point-to-point interfaces allow.RCCL provides practitioners with the ability to choose an allocator that is best-suited for e.g., in-kernel computations, while achieving high inter-APU communication bandwidth

\begin{comment}
\begin{tcolorbox}
\textbf{Observation 7}: Using RCCL as the communication interface provides a significant speedup in communication time over MPI.
\end{tcolorbox}
\end{comment}

{\iffalse
\begin{figure}[bt]
    \centering
    \includegraphics[width=\linewidth]{figures/clover_comm_speedup.pdf}
    \caption{Speedup in the communication part of the CloverLeaf HIP version. For each allocator, the results for MPI implementation are used as baseline. Higher is better. \todo{merge with Figure~\ref{fig:clover-comm-time}}}
    \label{fig:clover-comm-speedup}
\end{figure}
\fi}
{\iffalse
\begin{figure}
    \centering
    \includegraphics[width=\linewidth]{figures/clover_speedup_vs_alloc_xnackon_bm2048.pdf}
    \caption{Speedup in CloverLeaf communication part, for RCCL implementation, with MPI as baseline, with the three evaluated allocators.}
    \label{fig:clover-speedup}
\end{figure}
\fi}

% \subsection{AI?}
% \todo{(missing) some LLM training on four GPUs, which has been tested on AMD}
% \todo{Suggestion: inference: llama.cpp + model LLama 3.1 (405B*FP8 = 405GB = 4 MI300A); even possibility to change the allocator in GGML - main challenge due to the size of the workload, and is multi-GPU supported for HIP???. Also, we could even compare performance with one GH (fit the model in system memory), or with four GH (distribute model across GPUs).}

\section{Discussions}
\textbf{Allocators}. Depending on the data movement interface in use, the choice of allocator might affect the performance of data movement. In our experiments, hipMalloc is the only allocator for which the highest bandwidth of the Infinity Fabric interconnect could be reached consistently across all scenarios. In details, for hipMemcpy, using hipMalloc is required and MPI point-to-point routines are only able to achieve the maximum bandwidth for buffers allocated with hipMalloc. This is due to the MPI runtime delegating the copy operations performed on hipMalloc-allocated buffers to GPU hardware such as SDMA engines and GPU copy kernel, therefore achieving maximum bandwidth. For other choices of allocators, RCCL is the only programming interface that can utilize the full Infinity Fabric bandwidth, independent of the allocator, appearing as a solution to operate high-bandwidth data movement when the choice of allocator is constrained. Other factors, such as allocation time, which is higher with hipMalloc than with malloc, might be taken into account.

\begin{figure}[bt]
    \centering
    \includegraphics[width=\linewidth]{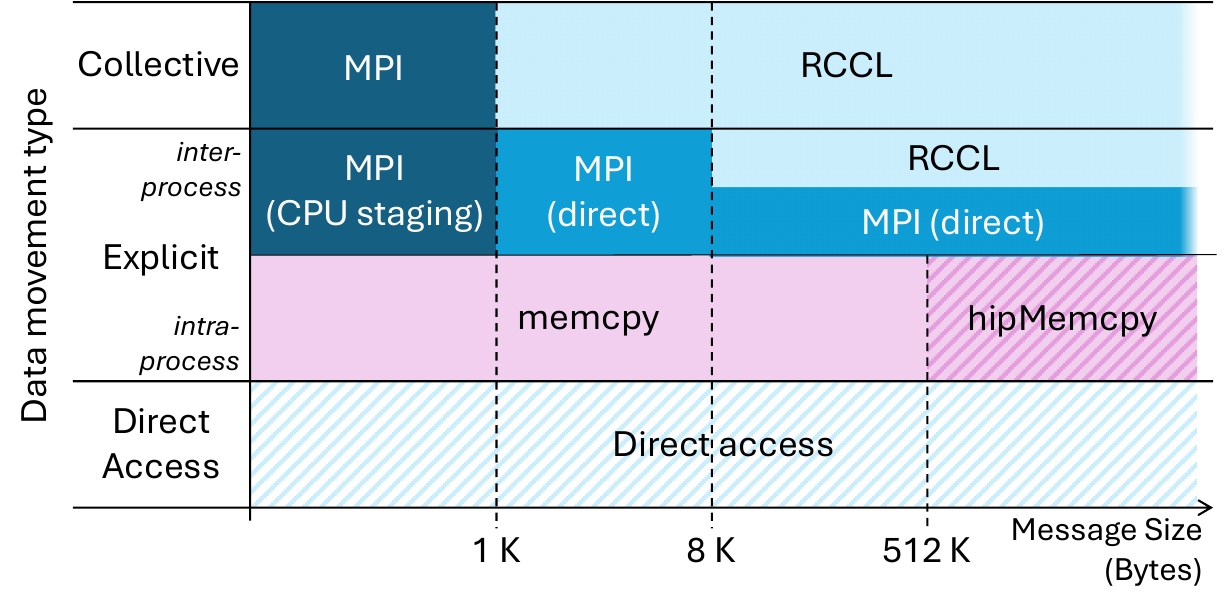}
    \caption{A summary of best-performing interface for inter-APU communication at various message sizes and data movement types, targeting to high bandwidth for explicit and direct access and low latency for collective operations. Assume buffers are allocated with hipMalloc.}
    \label{fig:summary_movement_if}
\end{figure}

\textbf{Programming Interfaces}. For each data movement scenario presented in the taxonomy Section~\ref{sec:taxonomy}, Figure~\ref{fig:summary_movement_if} presents the optimal interface depending on various message sizes. In general, for explicit and collective data movement at small messages, CPU-centric interfaces, namely memcpy and MPI with CPU-staging, provide the highest performance. This is a consequence of the high latency observed for GPU-centric interfaces, which is detrimental on small message sizes, which are typically latency-bound. However, those GPU-centric interfaces are able to leverage GPU hardware to perform data movement, namely SDMA engines and GPU copy kernels. They can therefore leverage the full Infinity Fabric bandwidth, making them suitable for larger message sizes. In contrast, this is not possible with memcpy and MPI with CPU-staging, due to those interfaces utilizing solely CPU resources to perform data movement.

\textbf{Communication Patterns}. Our evaluation of HPC applications focuses on two applications with explicit data movement, which rely on explicit inter-process communication routines, including point-to-point and collective operations. Other applications might have unpredictable communication patterns, where the extent of the accessed data is unknown at runtime, such as graph processing applications. On the AMD MI300A tightly-coupled system, those applications can benefit from direct data access from GPU kernel, which provides granular access to remotely-located data. We demonstrated that such access strategy achieves full utilization of the Infinity Fabric bandwidth, with approximately the twice latency of local access. 

\section{Related Works}
\textbf{AMD MI250X and MI300A.} Vijayaraghavan et al.~\cite{vijayaraghavan2017design} introduced the concept of APU integrated with CPU and GPU coupled in-package high-bandwidth memory for exascale computing, later applied on MI300A. Smith et al.~\cite{smith2024realizing} reported the technical details on the MI300A APU and key manufacturing insights. Porting and evaluation of HPC applications to one MI300A APU have been proposed by OpenMP's unified memory model~\cite{bertolli2024performance,tandon2024porting}. The previous generation of AMD GPUs, AMD MI250X, has previously been studied from various perspectives, including its Infinity Fabric interconnect~\cite{pearson2023interconnect} and Matrix Cores~\cite{schieffer2024rise}. Our work focuses on multi-MI300A node architecture, which is a key component of the latest leadership HPC systems.
 
% focus on the interconnect performance across on-node MI250X GPUs. They observed the improved performance of implicit access by mapped buffer rather than DMA engine control. Schieffer et al evaluate Matrix Cores on AMD MI250X GPUs, demonstrating the potential for the next-generation AMD GPU equipped with similar Matrix Cores architecture like MI300. 

\textbf{Evaluation of GPU-GPU Interconnects.}
GPU-GPU interconnects have been widely studied, including Nvidia's NVLink interconnect~\cite{li2018tartan}, and previous generations of Infinity Fabric interconnect, with AMD MI100 and MI250X GPUs~\cite{de2024exploring,khorassani2023high,pearson2018numa}.
De Sensi et al.~\cite{de2024exploring} evaluated GPU-GPU communication at large scale on three supercomputers, using both inter- and intra-node benchmarks. The similarity and difference on several interconnect characteristics have also been explored, in particular AMD MI250X based systems~\cite{siefert2023latency,de2024exploring}. Atchley et al.~\cite{atchley2023frontier} provided a large-scale evaluation of the Frontier supercomputer, including intra-node and inter-node evaluation. Khorassani et al.~\cite{khorassani2023high} evaluated Slingshot-interconnected nodes based on AMD MI100 GPUs. Hidayetoglu et al.~\cite{hidayetoglu2024commbench} focused on the multiple hierarchies in supercomputer interconnects. Schieffer et al.~\cite{schieffer2024understanding} characterized point-to-point and collective communication and memory allocation strategies on multi-GPUs MI250X-based supercomputers. In this work, we focus on the Infinity Fabric interconnected multi-MI300A compute nodes.

%Li et al.~\cite{li2018tartan} provide the Tartan benchmark suite to evaluate the interconnect between Nvidia GPUs. Pearson et al.~\cite{pearson2018numa} design a NUMA-aware data transfer benchmark suite for evaluating interconnect performance over various interfaces.
%Daniele et al.~\cite{de2024exploring} evaluate the performance of three supercomputer interconnects, including AMD MI250X nodes. They focused on the intra- and inter-node GPU communication on up to 4096 GPUs. Siefert et al.~\cite{siefert2023latency} focused on intra-node measurements of bandwidth and latency on a set of supercomputers, in particular on Frontier's AMD MI250X nodes. In Atchley et al.\cite{atchley2023frontier}, a large-scale characterization work of Frontiner included performance analysis of GPU interconnect for both intra-node and inter-node communication. Khorassani et al.~\cite{khorassani2023high} evaluated Slingshot-interconnected nodes that are based AMD MI100 GPUs. They focus only on MPI implementations and included Cray MPICH, OpenMPI, RCCL, and MVAPICH2 on early access cluster of the Frontier system. Hidayetoglu et al.~\cite{hidayetoglu2024commbench} provides a library with cross system backends including MPI, NCCL, RCCL for nodes based on Nvidia and AMD and Intel GPU networks.

\textbf{Multi-GPU Optimizations.}
Distributed multi-GPU systems are used ubiquitously on HPC systems and Data centers to accelerate a wide range of applications, including large language models, quantum computer simulations, and database query processing. In multi-GPU applications, understanding data movement patterns and bottlenecks is critical for performance. Such analysis was conducted on, e.g, Graph Neural Networks (GNN) applications~\cite{cai2021dgcl} and Convolutional Neural Networks (CNN)~\cite{tallent2018evaluating,shi2018performance}. To tackle the GPU-GPU communication bottleneck, several solutions have been proposed, including efficient workload partitioning using CUDA features~\cite{ocetkiewicz2024multi} and leveraging multiple path or CPU-GPU interconnects~\cite {lutz2020pump}. 
Young et al.~\cite{young2018combining} quantified the multi-GPU interconnect bottleneck with NUMA-aware software solutions like work scheduling, page placement, page migration, page replication, and caching remote data; and proposed co-design optimization strategies. Our work, especially the characterization results on multi-APU systems, provides a strong foundation for optimizing these applications and workloads on the emerging HPC systems and data centers. 
\section{Conclusions}
In this work, we evaluated inter-APU communication on Infinity Fabric on AMD MI300A systems. We quantified the peak hardware capacity and evaluated performance efficiency for various communication patterns, including CPU-GPU, point-to-point GPU-GPU, and GPU collectives. Our results quantified the impact of memory allocators and programming interfaces for data movement. Finally, we applied the optimization strategy on GPU-GPU communication in Quicksilver and CloverLeaf on four MI300A APUs, achieving a 2.15$\times$ speedup in communication.

%\section*{Acknowledgments}
%\todo{}
\begin{acks}
This work was performed under the auspices of the U.S. Department of Energy by Lawrence Livermore National Laboratory under Contract DE-AC52-07NA27344. LLNL-CONF-2004814. This research is supported by the Swedish Research Council (no. 2022.03062) and LLNL LDRD project 24-ERD-047. This work has received funding from the European High Performance Computing Joint Undertaking (JU) and Sweden, Finland, Germany, Greece, France, Slovenia, Spain, and Czech Republic under grant agreement No 101093261. 
\end{acks}

\printbibliography
\appendix

\end{document}